\documentclass[a4paper,fleqn,usenatbib]{mnras}

\usepackage[T1]{fontenc}
\usepackage{ae,aecompl}

\usepackage{graphicx}	
\usepackage{amsmath}	
\usepackage{amssymb}	

\title[High spatial resolution optical imaging of LkH$\alpha$ 262/263]{High spatial resolution optical imaging of the multiple T Tauri system LkH$\alpha$ 262/LkH$\alpha$ 263}

\author[S. Velasco et al.]
{S. Velasco,$^{1,2}$\thanks{E-mail: svelasco@iac.es}
	R. Rebolo,$^{1,2,3}$ A. Oscoz,$^{1,2}$ C. Mackay,$^{4}$ L. Labadie,$^5$  \newauthor A. P\'erez Garrido,$^6$ 
	J. Crass,$^{4,7}$ A. D\'iaz-S\'anchez,$^{6}$ B. Femen\'ia,$^8$ V. Gonz\'alez-Escalera,$^{1,2}$ \newauthor  D.L. King,$^4$ 
	R.L. L\'opez,$^{1,2}$    M. Puga,$^{1,2}$ L.F. Rodr\'iguez-Ramos$^{1,2}$ and J. Zuther$^5$ \\
	$^{1}$Instituto de Astrof\'isica de Canarias, c/V\'ia L\'actea s/n, La Laguna, E-38205, Spain\\
	$^{2}$Departamento de Astrof\'isica, Universidad de La Laguna, La Laguna, E-38200, Spain\\
	$^{3}$Consejo Superior de Investigaciones Cient\'ificas, Madrid, Spain\\
	$^{4}$Institute of Astronomy, University of Cambridge, Madingley Road, Cambridge CB3 0HA, United Kingdom\\
	$^{5}$I. Physikalsiches Institut, Universit\"{a}t zu K\"{o}ln, Z\"{u}lpicher Strasse 77, 50937 K\"{o}ln, Germany\\
	$^{6}$Departamento de F\'isica Aplicada, Universidad Polit\'ecnica de Cartagena, Cartagena, E-30202, Spain\\
	$^{7}$Department of Physics, University of Notre Dame, Notre Dame, IN, 46556, USA\\
	$^{8}$W. M. Keck Observatory, 65-1120 Mamalahoa Hwy., Kamuela, HI 96743, Hawaii, USA\\
}

\date{Accepted 2016 May 4. Received 2016 May 4; in original form 2015 July 17}

\pubyear{2015}

\begin{document}
\label{firstpage}
\pagerange{\pageref{firstpage}--\pageref{lastpage}}
\maketitle

\begin{abstract}
We report high spatial resolution \textit{i'} band imaging of the multiple T Tauri system LkH$\alpha$ 262/LkH$\alpha$ 263 obtained during the first commissioning period of the Adaptive Optics Lucky Imager (AOLI) at the 4.2 m William Herschel Telescope,  using its Lucky Imaging mode.  AOLI images have provided photometry for each of  the two components LkH$\alpha$ 263 A and B (0.41 arcsec separation) and marginal evidence for an unresolved binary or a disc  in  LkH$\alpha$ 262. The AOLI data combined with previously available and newly obtained optical and infrared imaging show that the three components of  LkH$\alpha$ 263 are co-moving, that there is orbital motion in the AB pair, and, remarkably, that  LkH$\alpha$ 262-263 is a common proper motion system  with less than 1 mas/yr relative motion. We argue that this is a likely five-component gravitationally bounded system.
According to BT-settl models the  mass of each of the five components is close to 0.4  M$_{\odot}$ and the age is in the range 1-2 Myr. The presence of discs in some of the components offers an  interesting opportunity to investigate the formation and evolution of  discs in the early stages of multiple very low-mass systems. In particular, we provide tentative evidence that the disc in 263C could be coplanar with the orbit of 263AB.
\end{abstract}





\section{Introduction}
\label{sec:intro}

The study of multiple systems is key to understanding the evolution of low-mass pre-main sequence (PMS) stars. Co-evality and a common metallicity are two important factors facilitating  the comparison of observational properties  with  evolutionary models \citep[e.g.][]{2012ApJ...747..103E}. As the physical properties of the various objects in a multiple system have to be explained for a given age and metallicity, a comparison with evolutionary tracks may lead to more stringent constraints  on masses and could test the consistency of model predictions for low-mass objects \citep{1998ApJ...495..385H}. Furthermore, the photometric and spectroscopic study of young systems allows us to address the question of star--disc interplay in an environment  dynamically constrained by the orbital motion of the pair \citep{1994ApJ...421..651A,2000MNRAS.314...33B}. Although there are indications that 
high-multiplicity systems are much more frequent among very young stars \citep[e.g.][]{2006A&A...459..909C}, until now, only a few of these young low-mass stellar systems have been confirmed.

LkH$\alpha$ 262 and LkH$\alpha$ 263, in the MBM12 cloud, are particularly interesting objects because they may be composed of four very young  M-type stars and some of the components are known to host discs \citep{2003ApJ...593L.101H}. The MBM12 molecular cloud was thought to be the nearest one to the Sun, at a distance of 58--90 pc \citep{2000A&A...357..681H}. However, further studies   by \citet{2001ApJ...560..287L}  have proposed a much greater distance at 275 pc, which we will assume in this paper.

LkH$\alpha$ 263 (02h 56m 8.433s, +$20^{\circ}$ 03' 38.63") is a triple T Tauri system with its primary and secondary stars tentatively classified as M-type \citep{2001ApJ...560..287L,2009A&A...497..379M}, and with a separation of 0.41 arcsec. The third fainter component C lies at 4 arcsec north-east from AB. This component C is an M0-type star hosting an optically thick edge-on disc, which was  discovered with near-IR adaptive optics observations \citep{2002ApJ...571L..51J}. LkH$\alpha$ 262 (02h 56m 8.00s, +$20^{\circ}$ 03' 24.2"), is another  T Tauri M0-type star \citep{ 2000A&A...353.1044H, 2001ApJ...550L.197J} 15 arcsec south-west of LkH$\alpha$ 263. The possibility of LkH$\alpha$ 262/263 being part of the same system was already discussed in \citet{2002A&A...394..949C}, but to our knowledge this membership has not been demonstrated yet. 

To investigate further the nature of LkH$\alpha$ 262/263 and, in particular, to study whether these two systems  are linked, we performed optical high spatial resolution observations during the first commissioning of the Adaptive Optics and Lucky Imager (AOLI) at the 4.2-m William Herschel Telescope (Roque de los Muchachos Observatory, La Palma). AOLI is a new instrument designed to combine  adaptive optics (AO) and lucky imaging (LI) techniques to reach close to diffraction limited imaging  in the visible range. AO is now an established technique for improving the spatial resolution of large ground-based telescopes  \citep[e.g.][]{2012ARA&A..50..305D, 1993ARA&A..31...13B} and offers diffraction-limited images in the near infrared (NIR). Unfortunately, by now there are few efficient AO systems in the visible \citep[see e.g.][]{2013ApJ...774...94C, 2005ApJ...629..592G}. The LI technique \citep{1978JOSA...68.1651F, 1964JOSA...54...52H} is  an alternative to AO to reach the diffraction limit. Images are taken at a very high speed in order to sample those intervals in which the atmospheric column through which the wavefront travels can be regarded as stable. If that fraction of the images with the best Strehl ratio are stacked in a shift-and-add process, the equivalent to a near-diffraction limited observation can be obtained in 2--4 m class telescopes \citep{2006A&A...446..739L}. The fraction of images selected for each target depends on the atmospheric conditions, as explained in \citet{1981JOSA...71.1138B}. Examples of LI instruments are LuckyCam \citep{2002A&A...387L..21T} and  FastCam \citep{2008SPIE.7014E..47O}. Early work on AO and LI can be found in \citet{2009ApJ...692..924L}, who used  the Palomar telescope infrared AO system with LuckyCam, and \citet{2011MNRAS.413.1524F}, who combined FastCam  with the WHT AO system. Near-diffraction limited images in the visible were obtained with both configurations at the two telescopes.  These advances  paved the way for AOLI, an instrument designed to combine an efficient AO system in the visible with a high-sensitivity LI camera.  

In this paper after a brief description of AOLI, we report on its first-light observations of the multiple T Tauri system  LkH$\alpha$ 262/263 at the WHT. We have conducted a photometric, astrometric, and imaging study of this remarkable system. 

\section{Description of the instrument}
\label{sec:instru}

\begin{figure*}
	\includegraphics[width=1\textwidth]{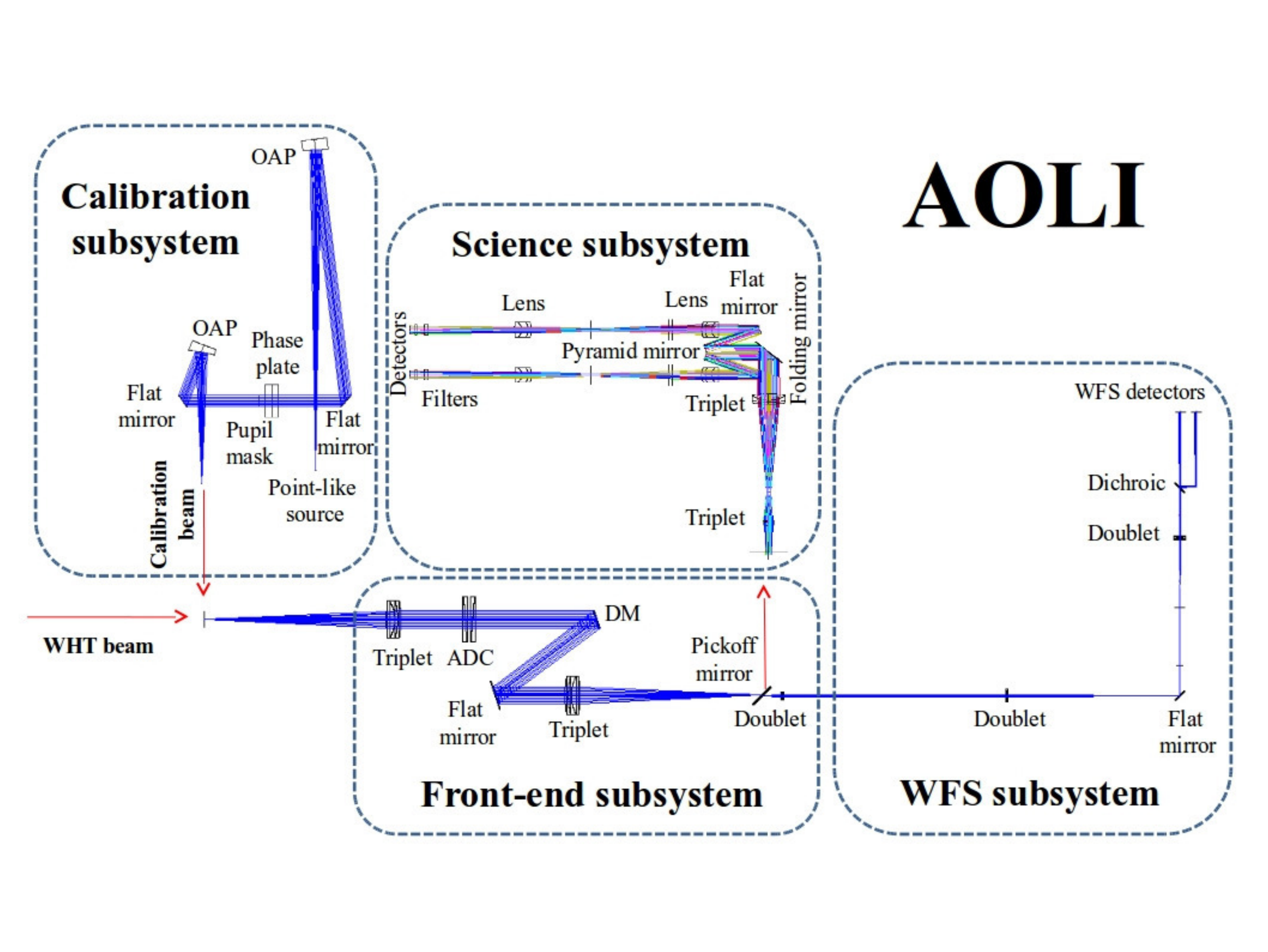}
	\centering
	\caption{The optical layout of AOLI remarking the calibration unit, the science camera, the wave front sensor arm and the path. The light enters AOLI, either from the telescope or from the calibration subsystem, from the bottom left-hand corner as indicated by the red arrows. Optical-path details and more layouts can be found in \citet{2013aoel.confE..11C} }
	\label{fig:aolifoto}
\end{figure*}

AOLI (Figure \ref{fig:aolifoto}) has been  developed by a consortium composed of the Instituto de Astrof\'isica de Canarias, the Institute of Astronomy at Cambridge University,  Universit\"{a}t zu K\"{o}ln, Universidad Polit\'ecnica de Cartagena, and Universidad de la Laguna. It is a visitor instrument at one of the WHT's Nasmyth platforms and it consists of seven major subsystems: front-end, mechanical support, calibration unit, science camera, AO unit, and control  and data analysis software. Complementary information and optical layouts can be found in \cite{2014SPIE.9147E..1TM,2012SPIE.8446E..21M}. 

The collected light, after passing through a derotator and a collimator lens, goes through an atmospheric dispersion corrector, which allows AOLI to operate at low elevations. Then, and after being reflected at the deformable mirror (DM) and a flat mirror, the beam is focused by a lens on to the pick-off mirror, whence light coming from the reference star is sent to the wavefront sensor (WFS). The remaining light is reflected by the pick-off mirror towards the science camera, which contains a set of lenses allowing a range of magnification scales giving a corresponding field of view from 36$\times$36 arcsec up to 120$\times$120 arcsec.

A pyramidal mirror architecture is placed at the entrance of the science camera system to re-image contiguous zones, each to a separate detector. Furthermore, the whole imaging structure can be moved in the  XYZ-axes for alignment, centring, and focusing purposes. The wide-field science camera is optimized for the 500 nm to 1 micron wavelength range with an array of four photon-counting, electron-multiplying, and back-illuminated E2V EMCCD detectors. The four 1024$\times$1024 pixels detectors, with very high quantum efficiency (peak over 95 per cent), are housed in a cryostat to improve their performance. Each has its own filter wheel allowing the use of a narrow-band filter for the science object with a broad-band filter for the reference star. These EMCCDs are able to offer images at 25 frames per second (fps) full frame. 

AOLI is designed to perform low-order corrections with faint stars ({\it I\/} $\sim$ 16--17). Two curvature WFS algorithms are being developed for AOLI: a non-linear \citep{2013MNRAS.429.2019A,2011SPIE.8149E..09M,2007SPIE.6691E..0GG} and a geometric one \citep{2013OptEn..52e6601F}. Any of them will enable the instrument to be used over a much wider part of the sky than with a Shack--Hartmann WFS. A calibration unit to simulate the telescope optics and the atmospheric  behaviour  is also included. It is used not only for calibration purposes but also to help WFS optimization. A description of this unit can be found in \citet{2014SPIE.9147E..7VP}. 

The acquisition software allows the possibility of processing the speckle images in real time and permits the observer to check a live-view of them during the observing night.

\section{Observations}
\label{sec:observations}

On 2013 September 24 and 25, AOLI was installed at the Ground-based High Resolution Imaging Laboratory (GHRIL) Nasmyth platform of the WHT for its first commissioning without its AO subsystem fully deployed. For the science observations presented in this paper only the LI technique was used.

Due to poor weather conditions during the run, we did not use the highest magnification science camera lens, preventing us to reach the diffraction limit for this telescope in the \textit{i' SDSS} band (38.6 mas).

\subsection{Plate scale and orientation}
The particular characteristics of globular clusters -- the presence of thousands of stars with different spectral type and great dynamic range -- make them laboratories for tesing and calibrating new LI instruments. We selected the M15 globular cluster for first light observations because it is compact and hosts a dense core   with previous astrometric measurements by the  {\it HST\/} \citep{2014arXiv1410.5820B, 2002AJ....124.3255V}. Moreover, its core has been studied by our group using LI with FastCam \citep{2012MNRAS.423.2260D}.  Although the AOLI observations were carried out under quite adverse weather conditions and with only one of the four science EMCCDs, the performance (linearity, plate scale, and dynamic range) of the instrument could be assessed using stars of the core of M15.

We selected and combined the best 10 per cent of 1029 images of the core of M15 to produce a final image in which we detected (at 3$\sigma$ level) stars up to $\sim$17 magnitude in the \textit{i' SDSS} band. For this task we used our own algorithm, as described in \cite{2012MNRAS.423.2260D}, searching for the local maxima by comparison of each pixel with its 8 neighbours. The location for each object in the image was determined as the centroid of a 3$\times$3 pixels box centred around each maximum. The typical position error for the centroids measured with our algorithm was 0.1 pixels.


To calculate the plate scale we cross-correlated the detected sources with the catalogue by \citet{1994AJ....107.1745Y}, measuring a plate scale of 54.9$\pm$0.3 mas/pixel. In this catalogue, done with {\it HST\/} data, precise positions and U, V and I photometric measurements are given for 1190 stars around M15's core. The distribution of these stars did not cover our whole image so we took a second correlation with \citet{2002AJ....124.3255V} also done with {\it HST\/}, obtaining a plate scale of 55.0$\pm$0.3 mas/pixel. In addition, along the night, we took some binaries of well known separation to check that no drifts were present.

This cross-correlation was done with the first coefficients for equations 1 and 2 from \citet{2003ApJ...595..187M}. The tilt and second order coefficients were found to be negligible, and thus the transformation of pixels to coordinates was almost linear. From this transform we got the orientation angle with an error of 0.3 degrees.

\subsection{Observations of the LkH$\alpha$ 262/263 T Tauri system}
\subsubsection{WHT observations}
The LkH$\alpha$ 262/263 system was observed on the night of September 24 with AOLI in the standard  {\it i'\/}  band (769.5 nm central wavelength, 137 nm full width at half-maximum). Both LkH$\alpha$ 262 and 263, were located within the FOV of the first 1024$\times$1024 EMCCD detector of the AOLI science camera. 

We obtained 4600 individual images with an exposure time of 50 ms, giving a total on-source observation time of 230 s. Using a standard LI algorithm \citep[see e.g.][]{2011A&A...526A.144L}, only 10 per cent of the images were selected for building the final image. In each frame both targets LkH$\alpha$ 262 and 263 were included. We chose the brightest pixel in the speckle of LkH$\alpha$ 262 for the shift-and-add algorithm as the contamination between LkH$\alpha$ 263 A and B due to their proximity renders them less adequate as LI reference stars. A very small fraction of individual frames exhibiting saturated pixels (due either to cosmic rays or spurious electronic events) was removed.

\begin{figure} 
	\includegraphics[width=0.4\textwidth]{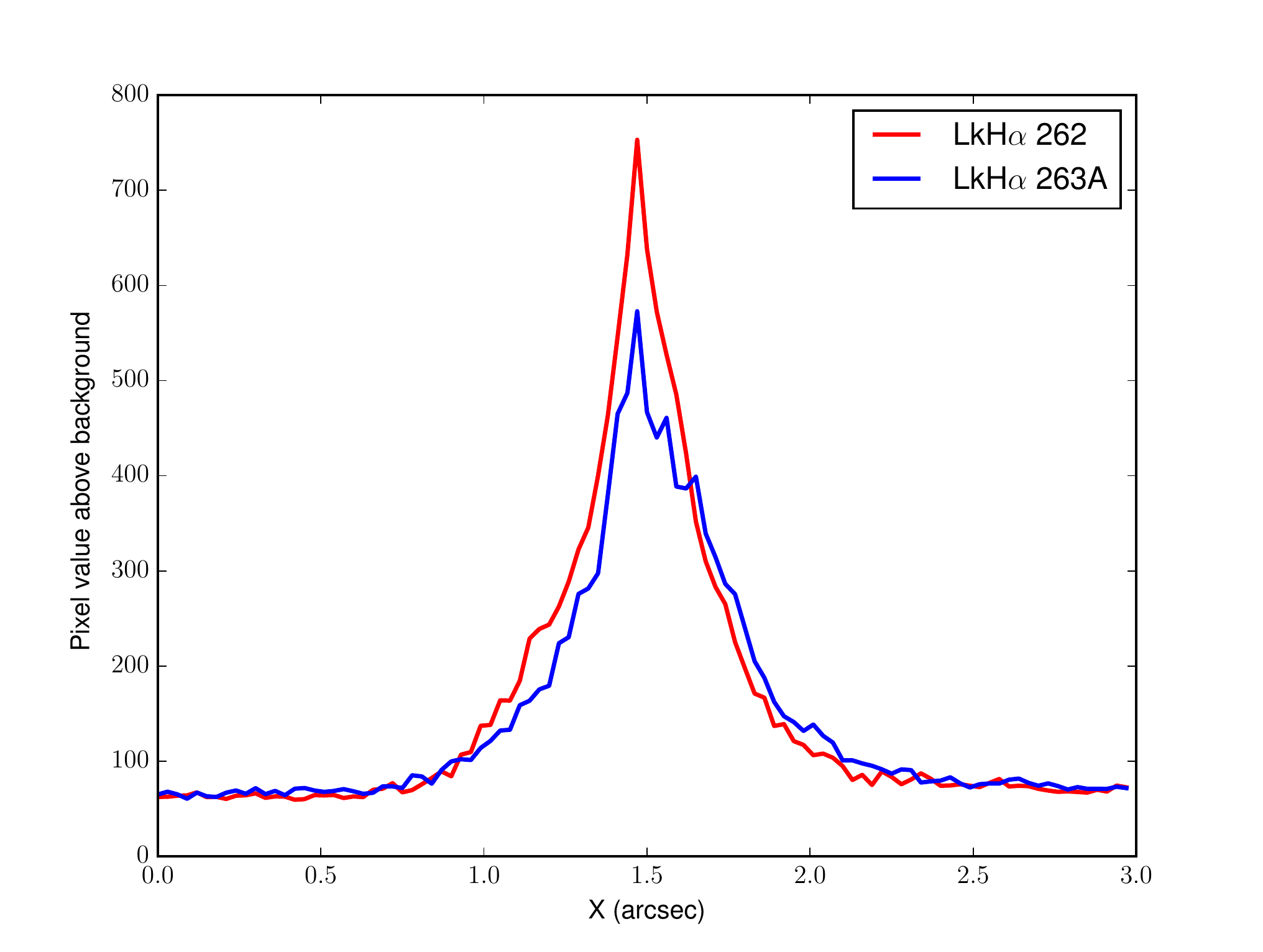}
	\centering
	\caption{The PSF of AOLI as derived from the best 10 per cent images of LkH$\alpha$ 262 (red) and LkH$\alpha$ 263A (blue) in the {\it i'\/} band. No AO subsystem was used in this case. Both profiles are obtained using  LkH$\alpha$ 262 as the reference star for the lucky imaging algorithm.}
	\label{fig:stf}
\end{figure}

Owing to the short time exposures and the shift-and-add algorithm at sub-pixel scales used in the LI-processed images, the point spread function (PSF) shows a classical Moffat shape presenting a narrow core and an extended halo. Despite poor seeing conditions (worse than 2 arcsec) we could obtain an FWHM for the best PSF of 0.15$\pm$0.01 arcsec (see Figure \ref{fig:stf}). A fully working AOLI system should deliver images of $\sim$0.04$\arcsec$ FWHM, close to the diffraction limit of the 4.2-m WHT at the wavelength of the observations.

\subsubsection{NOT observations}

Since the commissioning nights were not photometric, no standard star was observed during the run at the WHT. For calibration, we performed observations of the LkH$\alpha$ 262/263  system and several standard stars in the same \textit{i'} band with ALFOSC at the Nordic Optical Telescope (NOT, Observatorio del Roque de los Muchachos, La Palma, Spain) on 2014 September 15. Weather conditions during the NOT observations were photometric, with the average seeing ranging from 0.5 to 1 arcsec. This optical camera consists of a $2k\times2k$ CCD detector with a 0.187 arcsec/pixel plate scale, providing a field of view (FOV) of $6.4\times6.4$ arcmin$^2$. 

Routines within the IRAF\footnote{ IRAF is distributed by National Optical Astronomy Observatories, which is operated by the Association of Universities for Research in Astronomy, Inc., under contract to the National Science Foundation.} environment were used to reduce raw data, and we used the IDL\footnote{IDL (Interactive Data Language) is a product of ITT Visual Information Solutions.} CCDPHOT package \citep{1996ASPC..101..135B} to perform aperture photometry. These observations provided integrated photometry for LkH$\alpha$ 262 and LkH$\alpha$ 263AB as unresolved sources.

\subsubsection{TCS and IAC80 observations}

With the aim of obtaining astrometric measurements, additional observations were carried out with the IAC80 telescope and the Telescopio Carlos S\'anchez (TCS), of the Observatorio del Teide, in Tenerife. The IAC80 images were taken in the \textit{i'} band with the CAMELOT instrument \citep{2013ApJ...779..144O}, consisting of a 2k$\times$2k E2V CCD detector with a 0.304 arcsec/pixel plate scale, providing an FOV of $10.4\times10.4$ arcmin$^2$.  Meanwhile, the TCS observations were performed in the $J$ band with the CAIN infrared camera \citep{2006A&A...453..371C}. CAIN consists of a 256$\times$256 HgCdTe detector providing an FOV of 1.7 square arcmin with a corresponding 0.39 arcsec/pixel plate scale.

\subsubsection{Archive data}

In the analysis of this system we have used some previous archival observations from different instruments and telescopes, including:

- Hokupaa at Gemini North telescope, images taken by \citet{2002ApJ...571L..51J} on December 25th 2000 in the {\it J, H' and $K_{\rm s}$} bands. The values used here for this dataset have been taken from \citet{2002ApJ...571L..51J}, while the reduced images showing LkH$\alpha$ 263ABC were provided as private communication. These images do not show both LkH$\alpha$ 262 and LkH$\alpha$ 263ABC in the same FoV.

- Aobir at Canada-France-Hawaii telescope (CFHT), images in the {\it J \/} band obtained by \cite{2003ApJ...589..410S} on December 1st 2001. The plate scale used was 34.8$\pm$0.1 mas/pixel.

- SDSS images in the {\it i', r' and z'\/} bands obtained on September 21st 2004. The astrometric solution given on the reduced images was used to determine the projected separation of our targets. 

- IRCS at Subaru telescope, consisting of a series of images in the {\it H\/} band obtained on September 28th 2004 (unpublished). We used the given astrometric solution with a nominal plate scale of 20$\pm$1 mas/pixel.

- NACO at the Very Large Telescope (VLT) in the {\it L'\/} band obtained by S. Correia on October 7th 2004 (unpublished). This diffraction limited image is shown on Figure \ref{fig:comparanaco}. The plate scale used was 27.1$\pm$0.1 mas/pixel

- The ACS WFC camera at {\it HST\/}, consisting of two images in the {\it FW555\/} and {\it FW851\/} bands obtained on September 11th 2005 (unpublished). Although the objects LkH$\alpha$ 262 and 263AB are saturated in these images, the angular separation between them could be measured as described by \cite{2011PASP..123.1290R} using the astrometric solution given with a plate scale of 49.6$\pm$0.1 and 49.5$\pm$0.1 mas/pixel in the x and y axis respectively.

- LIRIS at  William Herschel Telescope (WHT) in the {\it hc\/} band taken as a field reference for spectroscopic studies on December 7th 2009 by B. Montesinos (unpublished). Due to the original purpose of these images there was not any photometric or astrometric reference, we used the given plate scale of 250$\pm$1 mas/pixel to determine the projected separation of our targets.

\section{Results and discussion}
\label{sec:results}

\subsection{Photometry}

Our initial goal was the photometric characterization in the optical of the three components of  LkH$\alpha$ 263. We used LkH$\alpha$ 262 as the LI reference target. We used routines from the DAOPHOT package to perform aperture photometry of 262 and 263C since it can be used for single stars and extended objects. We chose an aperture radius of 2.5 times the size of the FWHM. 

As components A and B are too close together, we used the values of their peaks to weight their contribution to the integrated photometry of LkH$\alpha$ 263AB. This procedure proved to give good results on LI produced images owing to the very narrow PSF shape, as shown in \citet{2011A&A...526A.144L}, the peak value method provides better accuracy than aperture photometry for very close objects. With the flux ratios obtained with AOLI and the calibrated absolute photometry obtained with ALFOSC we could determine the magnitude for each component and their relative photometry with a small error.

LkH$\alpha$ 262 is a known variable source, measured in the 2MASS All-Sky Catalog of Point Sources by \citet{2003yCat.2246....0C}. A maximum magnitude value of \textit{I}=12.51 and variation of 1.14 mag over a period of 1.27 days has been reported by \citet{2011IBVS.5969....1K}. Thus, our absolute calibration has to be taken with caution. However, the relative photometry of the various components LkH$\alpha$ 263A, B and C is much more reliable. In Table \ref{table:phot} we list the \textit{i'}-band photometry values obtained. Our \textit{i' SDSS} data show that stars LkH$\alpha$ 263 A and B are about five magnitudes brighter than component C. 

We also obtained photometry of the full system in the NIR to obtain the {i'-J} colour for each  component. The LkH$\alpha$ 263 system had been observed by \citet{2002ApJ...571L..51J} in the \textit{J}, \textit{H} and \textit{K$_{s}$} infrared bands using AO at the Gemini North telescope. We measured LkH$\alpha$ 262 in the \textit{J} band with CAIN at the TCS and determined the \textit{i'-J} value for the LkH$\alpha$ 262/263 components (see Table \ref{table:phot}). We derived the spectral type of our object by comparing with the \textit{i'-J} colours in table 3 of \citet{2002AJ....123.3409H}. For LkH$\alpha$ 262 we determined \textit{i'-J}=2.0$\pm$0.1, consistent with an M0/M1 star. Previous works have shown that LkH$\alpha$ 262 is a T Tauri star of M0 spectral type \citep{2001ApJ...550L.197J, 2000A&A...353.1044H}. 

As LkH$\alpha$ 263 A and B are so close, they have not been spectroscopically resolved yet, so that there is a greater uncertainty in their spectral classification, which ranges in the literature from M2 to M4, see \citet{2009A&A...497..379M} and \citet{2001ApJ...560..287L}. However, the measured colour suggests an M0- and M2-type spectral class for each of these components, respectively.

For component C we found a colour \textit{i'-J} = 1.12$\pm$0.25, which  would correspond to a K-type star, inconsistent with the actual spectral classification as M0 \citep{2002ApJ...571L..51J}. This discrepancy between the observed colour and the spectral class for component C probably resides in the presence of an edge-on disc in this object. It is interesting to compare the colours of LkH$\alpha$ 263C with those of well-known edge-on disc systems, such as HK Tau C or HV Tau C. \cite{1998ApJ...502L..65S} derive for HK Tau C a colour \textit{I-J} = 2.4$\pm$0.1 based on \textit{HST} optical images in the \textit{F814W} filter and ground-based CFHT infrared images obtained one month apart. For HV Tau C we use the \textit{F814W} optical and the CFHT/PUEO infrared magnitudes by \cite{2003ApJ...589..410S} about one year apart to derive a colour of \textit{I-J} = 0.85$\pm$0.1.

We list in Table \ref{table:discs} the colours and spectral classes for some T Tauri stars hosting edge-on discs. While variability effects need certainly to be considered at optical and NIR wavelengths (see \citet{2010ApJ...712..112D} for HV Tau C) for such young systems. We observe that our \textit{i'-J} colour measurement lies between those found for the HK Tau C and HV Tau C systems, and therefore not unusual (note that only an instrumental magnitude difference less than 0.06 mag is expected in the final colours owing to the \textit{i'} SDSS and \textit{F814W} filters having different shapes). Although an analysis of the disc physics is beyond the scope of this paper, a possible explanation for such a blue \textit{i'-J} colour in LkH$\alpha$ 263C is efficient dust scattering at short wavelengths, which then connects to more fundamental properties, as the characteristic size of the disc surface dust grains  \citep{2013A&A...549A.112M}.

\begin{table*}
	
		\caption{Photometric data and spectral classification of the LkH$\alpha$ 262-3 system.}
		\label{table:phot}
		\begin{tabular}{ccccccc}
			\hline
			Object      & \multicolumn{0}{c}{\textit{i'}} &  \textit{J}      & \textit{H}      & \textit{$K_{s}$} &\textit{i'-J} & Spectral type  \\ 
			\hline
			\hline
			LkH$\alpha$ 262 &  12.5$\pm$0.1$^{a}$ & 10.5$\pm$0.1$^{c}$ & 10.3$\pm$0.1$^{d}$ &  9.6$\pm$0.1$^{d}$ &    2.0$\pm$0.1    &  $\sim$ M0/M1$^{f}$ \\
			LkH$\alpha$ 263AB & 12.6$\pm$0.1 $^{a}$ &  &  & \\
			LkH$\alpha$ 263A & 13.4$\pm$0.1 $^{b}$    &  11.52$^{e}$  & 10.64$^{e}$       &  10.21$^{e}$ & 1.91$\pm$0.10     &  $\sim$ M0-M1$^{g}$ \\
			LkH$\alpha$ 263B &  13.2$\pm$0.1$^{b}$  &  11.25$^{e}$   & 10.51$^{e}$       &  10.34$^{e}$ & 1.95$\pm$0.10    &  $\sim$ M1-M2$^{g}$ \\
			LkH$\alpha$ 263C &  17.6$\pm$0.2$^{b}$   &  16.5$\pm$0.15$^{e}$    &   16.0$\pm$0.15$^{e}$    & 16.1$\pm$0.3$^{e}$ & 1.12$\pm$0.25    &   $\sim$ M0$^{h}$ \\
			\hline
		\end{tabular}
		\medskip
		\small
		\\
		$^{a}$Photometry obtained with  ALFOSC (NOT) on September 2014;  $^{b}$From delta magnitude photometry measured with AOLI on September 2013 and applied to ALFOSC photometry;  $^{c}$Photometry obtained with CAIN at TCS on November 2014;  $^{d}$From 2MASS \citep{2006AJ....131.1163S}; $^{e}$Photometric values from  \citet{2002ApJ...571L..51J} with Gemini-N telescope using AO on December 2000; $^{f}$AS Given by \citet{2009A&A...497..379M}; $^{g}$Based on table 3 from \citet{2002AJ....123.3409H} from \textit{i'-J} values; $^{h}$Given by \citet{2002ApJ...571L..51J}.

\end{table*}

\begin{table*}
	
		\caption{Colour and spectral classification of edge-on discs in T Tauri systems.}
		\label{table:discs}
		\begin{tabular}{ccccccc}
			\hline
			Object      & \multicolumn{0}{c}{\textit{I/i'}} &  \textit{J}   &\textit{I-J} & Spectral type  \\ 
			\hline
			\hline
			HK Tau C &  15.90$\pm$0.1$^{a}$ & 13.5$\pm$0.1$^{a}$ &    2.4$\pm$0.1    &  M2$^{b}$ \\
			HV Tau C &  15.50$\pm$0.1$^{c}$ & 14.66$\pm$0.1$^{d}$ &   0.85$\pm$0.1    &  M0$^{d}$ \\
			LkH$\alpha$ 263 C &  17.6$\pm$0.2$^{e}$ & 16.5$\pm$0.15$^{f}$ &    1.12$\pm$0.25    &  M0$^{f}$ \\
				\hline
		\end{tabular}
		\medskip
		\small
		\\
		$^{a}$\cite{1998ApJ...502L..65S};  $^{b}$\cite{1998A&A...339..113M};  $^{c}$\cite{2003ApJ...589..410S}; $^{d}$Monin 2000;$^{d}$\cite{1998A&A...338..122W}; $^{e}$From this paper (\textit{i'} band);  $^{f}$\citet{2002ApJ...571L..51J}. 
	
\end{table*}

\subsection{Astrometry}

We have measured the PSF centroids and determined positions for all the objects in the LkH$\alpha$ 262/263 system. Using our plate calibration, we determined the distances and angles between all the components detected in the final AOLI image.
\subsubsection{Evidence for LkH$\alpha$ 263 as a bounded system}
\begin{figure}
	\includegraphics[width=0.5\textwidth]{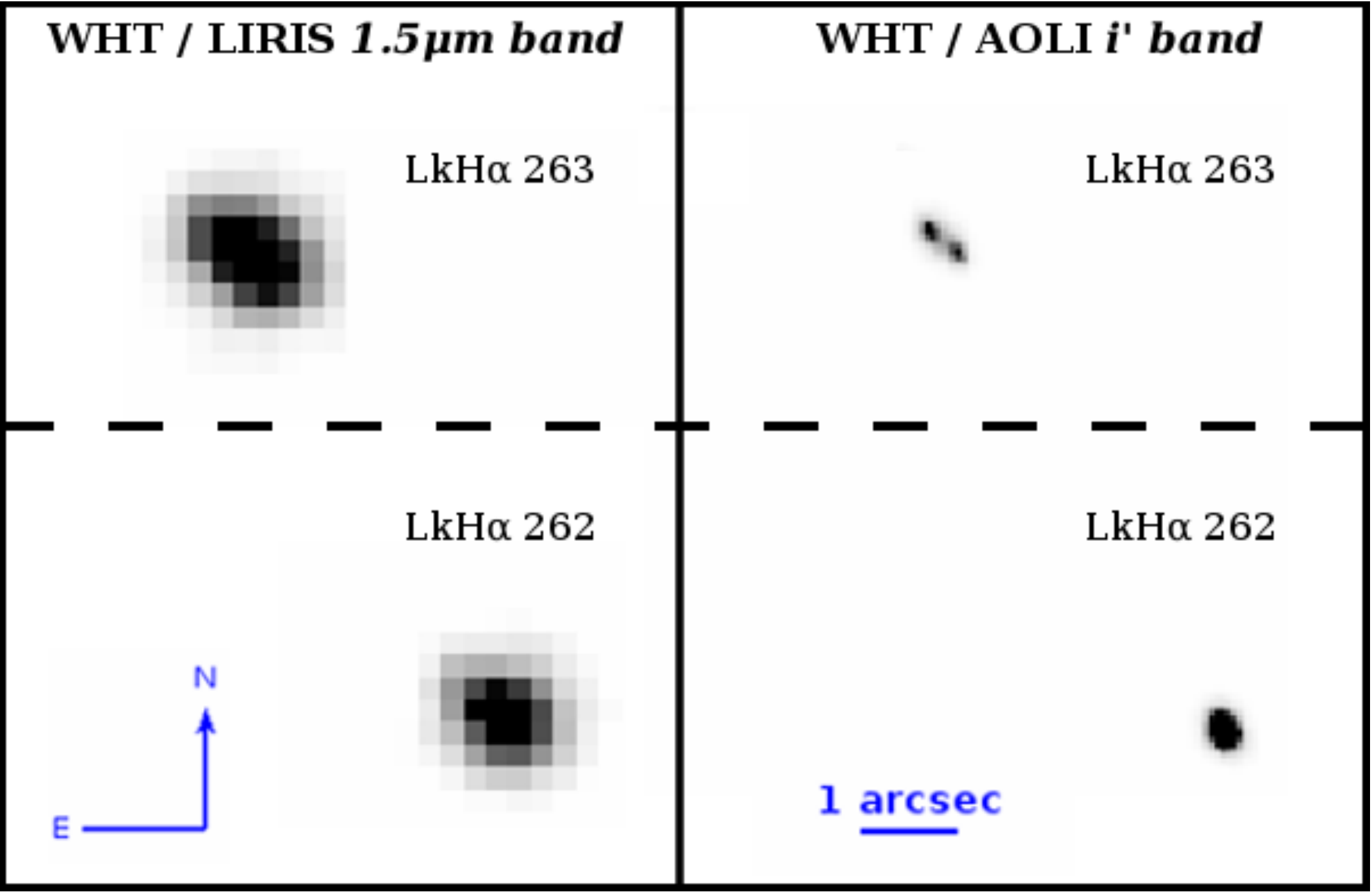}
	\centering
	\caption{Left: LkH$\alpha$ 262 (south) and LkH$\alpha$ 263 (north) image taken with LIRIS at the WHT by PI B. Montesinos in 2009 in the 1.5 $\mu$ band (5 seconds integration time). Right: The same field as observed with AOLI. In this image, 10 per cent out of 4600 individual 50 ms frames  were selected for lucky imaging processing. Components A and B of LkH$\alpha$ 263 are clearly resolved in the AOLI image. NOTE: Zig-zag line indicates that the empty space between LkH$\alpha$ 262 and 263 has been omitted. Scale has been preserved. }
	\label{fig:263ab}
\end{figure}

Comparison with previous archival  observations from the {\it HST\/}, LIRIS at the WHT, and with CAMELOT at the IAC80 (this study) shows that the position angle between LkH$\alpha$ 262 and 263 is stable; thus, a similar stability has also been assumed for the AOLI images. We did not use the derotator in our AOLI observations so the angle between LkH$\alpha$ 262 and 263 was taken as a fixed reference to rotate the individual images. The LIRIS and AOLI images, both taken with the WHT, can be compared in Figure \ref{fig:263ab}.

As shown in Table \ref{table:astrometry}, over a 13 year period, components B and C in LkH$\alpha$ 263 have remained co-moving with component A with a differential motion of less than 1 mas/yr between A and B, and less than 10 mas/yr for A and C.    

\begin{figure}
	\includegraphics[width=0.5\textwidth]{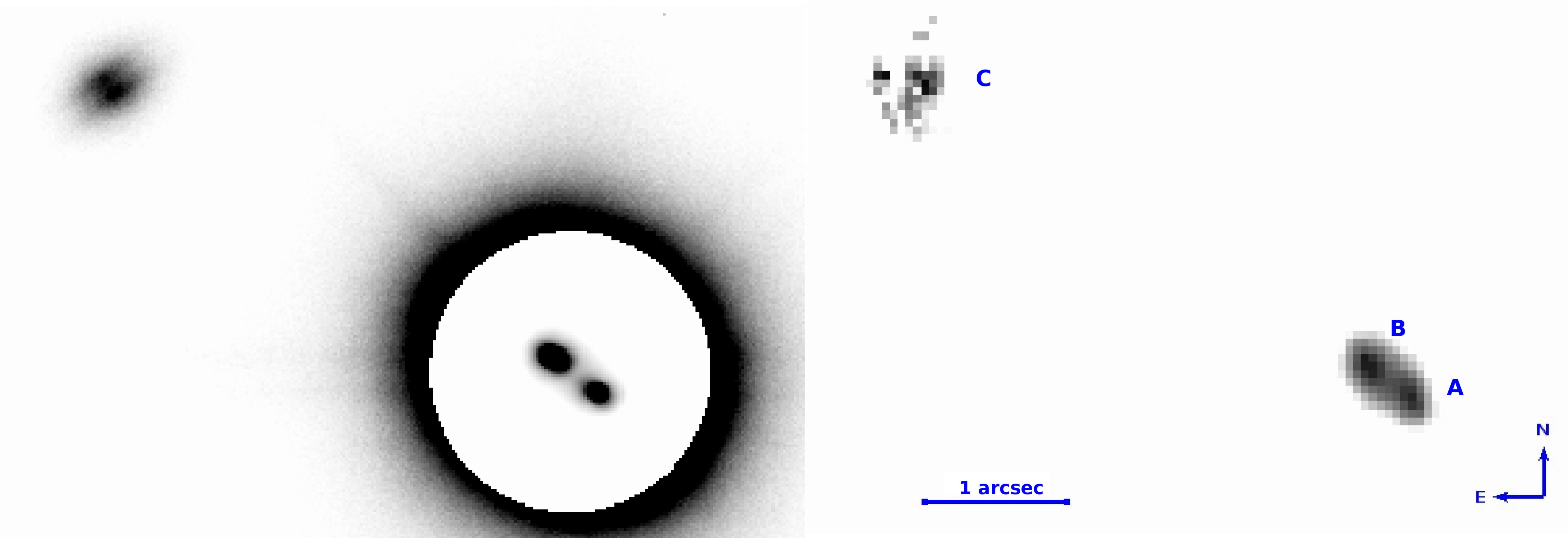}
	\centering
	\caption{The LkH$\alpha$ 263 system. Left: image obtained by Jayawardhana et al. in 2000 (private communication). LkH$\alpha$ 263C is visible as a double disc. Right: image obtained with AOLI in 2013, LkH$\alpha$ 263C is visible together with  LkH$\alpha$ 263AB. Both images are composites, the flux of component C has been enhanced while A and B have been lowered below the saturation level in order to show all three components in both epochs for a better comparison.}
	\label{fig:263compara}
\end{figure}

 In Table \ref{table:astrometry}, a comparison with previous data in the infrared is given and shows a 4$\sigma$ detection of orbital motion of the AB pair, amounting 1.5 degrees over a period of 13 years. In Figure \ref{fig:263compara} it is possible to discern visually this small rotation suffered by component B with respect to A between years 2000 \citep{2003AJ....126.2009B} and 2013 (AOLI).

\begin{table*}
	
	\begin{minipage}{150 mm}
		\caption{Astrometric measurements for the LkH$\alpha$ 263 system.}
		\label{table:astrometry}
		\centering
		\begin{tabular}{ccccc}
			Component & Obs. date & Instrument & Separation  & Position  Angle   \\
			&   & (band) & \( [arcsec] \) & \([deg ]\)  \\
			\hline
			\hline
			A-B	    & 22/02/2000 $^{a}$     &   Keck (\textit{J,H,K'}) & 0.416$\pm$0.003    & 52.67$\pm$0.05 $^{*}$       \\
			& 25/12/2000 $^{b}$    &   Gemini (\textit{K$_{s}$}) & 0.415$\pm$0.004    & 51.9$\pm$0.1            \\
			& 01/12/2001 $^{c}$    &   CFHT (\textit{J,H,K$_{s}$})& 0.416$\pm$0.003    & 54$\pm$1           \\
			& 14/11/2002 $^{d}$    &   VLT (\textit{Br$_{\gamma}$})& 0.414$\pm$0.001    & 52.0$\pm$0.5           \\
			& 28/09/2004 $^{e}$    &   Subaru (\textit{H})& 0.414$\pm$0.005    & 51.6$\pm$0.2           \\
			& 24/09/2013 $^{e}$    &   AOLI (\textit{i'})& 0.408$\pm$0.009    & 51.1$\pm$0.4           \\
			A-C       & 25/12/2000 $^{b}$     &  Gemini (\textit{K$_{s}$})  & 4.12$\pm$0.02     & 58.3$\pm$0.2         \\
				& 01/12/2001 $^{c}$    &   CFHT (\textit{J,H,K$_{s}$})& 4.10$\pm$0.05    & 61$\pm$1           \\
			& 28/09/2004  $^{e}$    &   Subaru (\textit{H})   & 4.04$\pm$0.03    & 57.7$\pm$0.4           \\
			& 24/09/2013   $^{e}$   &   AOLI (\textit{i'})   & 3.99$\pm$0.03    & 57.3$\pm$0.6           \\
			\hline
		\end{tabular}
		\small
		\\
		$^{a}$\cite{2003AJ....126.2009B}; $^{*}$PA given as 232.67$\pm$0.05 due to swap of A and B components;  $^{b}$\cite{2002ApJ...571L..51J}; $^{c}$\cite{2002A&A...394..949C}; $^{d}$\cite{2006A&A...459..909C}; $^{e}$From this paper; NOTE - The positions and angles of components B and C are measured relative to component A.
	\end{minipage}
\end{table*}

\subsubsection{Constraining the quadruple nature of the system with proper motions}

\citet{2005A&A...438..769D} set upper limits for the relative motion of LkH$\alpha$ 263AB (centroid) and LkH$\alpha$ 262 at less than 8 mas/yr (-3$\pm$6 mas/yr in RA and +5$\pm$6 mas/yr in DEC).

\begin{figure*}
	
		\includegraphics[width=1\textwidth]{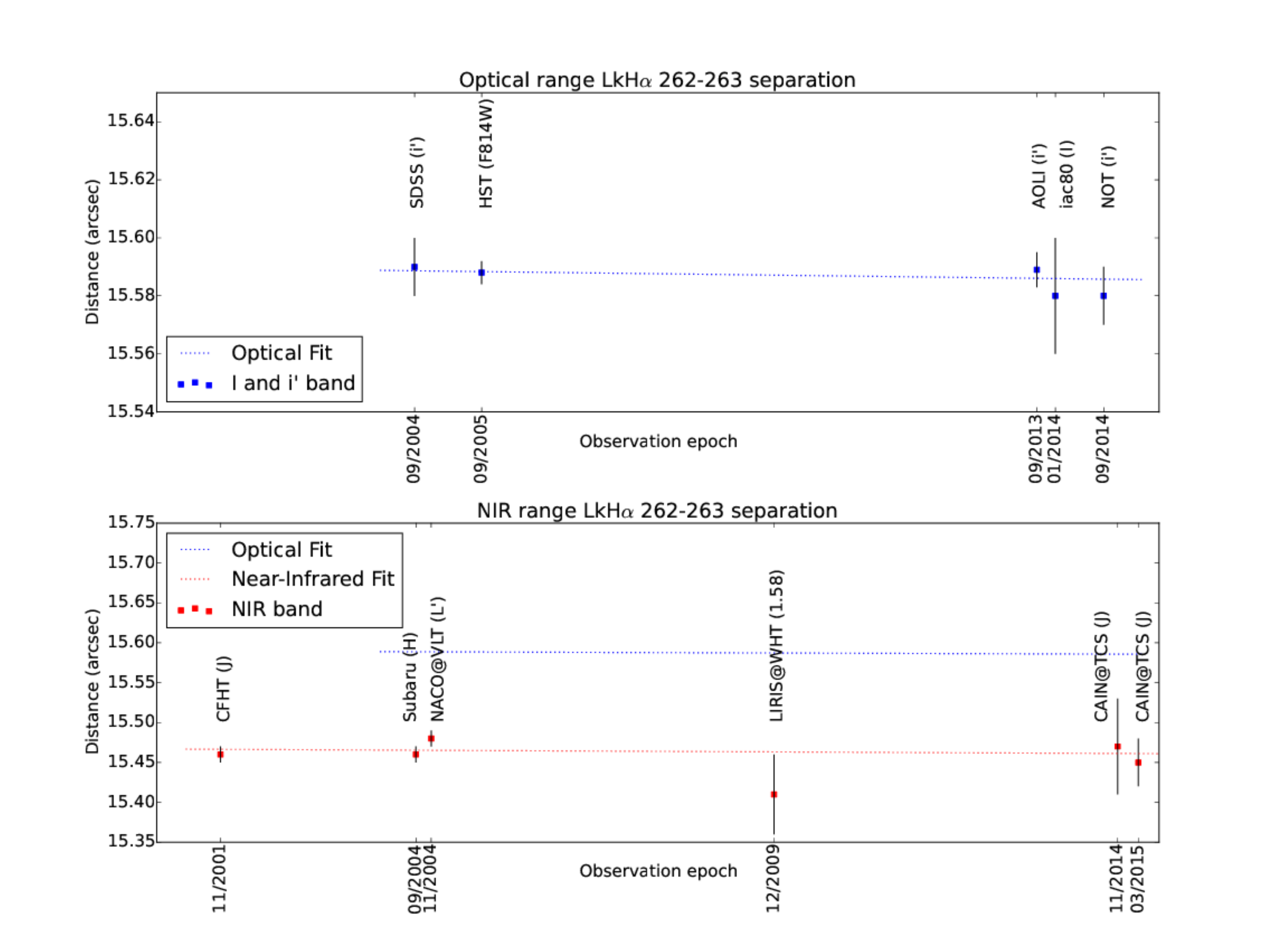}
		\centering
		\caption{Top: The separation of the LkH$\alpha$ 262/263 system measured with various instruments and at different epochs in the optical range. A trend line (dotted blue) has been added. The \textit{HST} stars are saturated; however, their centroids could be measured, thus proving the distance to be compatible with that obtained with the other instruments.
			Bottom: The separation of the LkH$\alpha$ 262/263 system measured with different instruments and epochs in the NIR. A dotted red trend line has been added. The dotted blue line represents the trend line for optical data. The Python curvefit routine has been used to take into account the weighting errors.}
		\label{fig:pm}
	
\end{figure*}

With AOLI  we measured a distance of 15.589$\pm$0.006 arcsec between the centroids of LkH$\alpha$ 263AB and LkH$\alpha$ 262. We list in Table \ref{table:astrometry2} the projected separation for these two systems using images available in both the optical and infrared. We find the separation to be very stable in time in both optical and NIR measurements, as can be seen in Figure \ref{fig:pm}, which includes new data taken by us. The trend lines in Figure\ref{fig:pm} are consistent with a zero slope to within the error bars for both the optical and infrared: slope of -0.29$\pm$0.35 mas/yr and -0.4$\pm$2.0 mas/yr, respectively for the optical and infrared measurements. 

Our measurements provide evidence for a common proper motion of LkH$\alpha$ 262 and 263.  The upper limit on the differential motion imposed by the optical observations strongly suggests a gravitational bound between  LkH$\alpha$ 262 and 263. In an area of 20$\times$20 square degrees in the direction of the LkH$\alpha$ 262/263 system, the stars at distances of 250--300 pc have proper motions 10 times larger \citep{1997A&A...323L..49P}. Moreover, less than 5 per cent of the stars within this distance range have a proper motion lower than 5 mas/yr.

Besides this, the density distribution of the MBM12 cloud at the region of these objects is only about five  members per square degree. Thus, the typical separation of cluod members is more than 10 times larger than the separations between   LkH$\alpha$ 262 and LkH$\alpha$263, potential gravitational link.

\begin{table*}
	\begin{minipage}{100mm}
		\caption{Angular separation measurements of the LkH$\alpha$ 262-263 system.}
		\label{table:astrometry2}
		\begin{tabular}{lcccc}
			Instrument@Telescope &  Observation date & Band & Separation     \\
			&  \(\) & & \( [arcsec] \)  \\
			\hline
			\hline

			Hokupaa@Gemini-N & 25/12/2000 		 		& \it{J - H'}   & 15.50           \\
			aobir@CFHT   	&  1/12/2001	& \it{J}   & 15.46$\pm$0.01           \\
			SDSS       & 21/09/2004 	& \it{i'}   & 15.59$\pm$0.01           \\
			&  				 	& \it{z'}   & 15.47$\pm$0.01          \\
			&  				 	& \it{r'}   & 15.57$\pm$0.01           \\
			&  				 	& \it{g'}   & 15.63$\pm$0.01           \\       
			IRCS@Subaru     & 28/09/2004     & \it{H}   & 15.46$\pm$0.01          \\
			NACO@VLT     & 07/10/2004     & \it{L'}   & 15.48$\pm$0.01          \\
			ACS@HST    &  11/09/2005 	& \it{FW555}   & 15.595$\pm$0.004           \\
			&  				 	& \it{FW814}   & 15.588$\pm$0.004          \\
			LIRIS@WHT      &  07/12/2009	 & \it{hc}   & 15.41$\pm$0.05           \\
			AOLI@WHT     & 	24/09/2013  	 & \it{i'}   & 15.589$\pm$0.006           \\
			CAMELOT@IAC80   &  26/01/2014		 & \it{I}   & 15.58$\pm$0.02          \\
			ALFOSC@NOT   &  15/09/2014		 & \it{i'}   & 15.58$\pm$0.01          \\
			CAIN@TCS   &  24/11/2014		 & \it{J}   & 15.47$\pm$0.06          \\
			CAIN@TCS   &  05/03/2015		 & \it{J}   & 15.45$\pm$0.03         \\                   
			
			\hline
		\end{tabular}
	\end{minipage} 
\end{table*}

As can be directly noticed in Figure \ref{fig:pm}, the mean value of the trend in the optical band is 15.59$\pm$0.01, whereas in the infrared it is 15.46$\pm$0.01. We ascribe the discrepancies between the optical and infrared measurements (of order 130 mas) to the likely presence of a barely resolved companion and/or a disc in LkH$\alpha$ 262,  as we discuss below.

One possible explanation would be the existence of a fainter and cooler companion to LkH$\alpha$ 262 located NE of the primary. AOLI's LkH$\alpha$ 262 \textit{i'}-band data show an elongation of the PSF in this direction, which could suggest the existence of an unresolved faint companion, at an angular separation below the 0.15 arcsec angular resolution achieved by the current AOLI observation. Its relative contribution respect to the primary is expected to be greater in the infrared than in the optical, thus altering the position of the centroid between these two bands.

\begin{figure}
	\centering
	\includegraphics[width=1\linewidth]{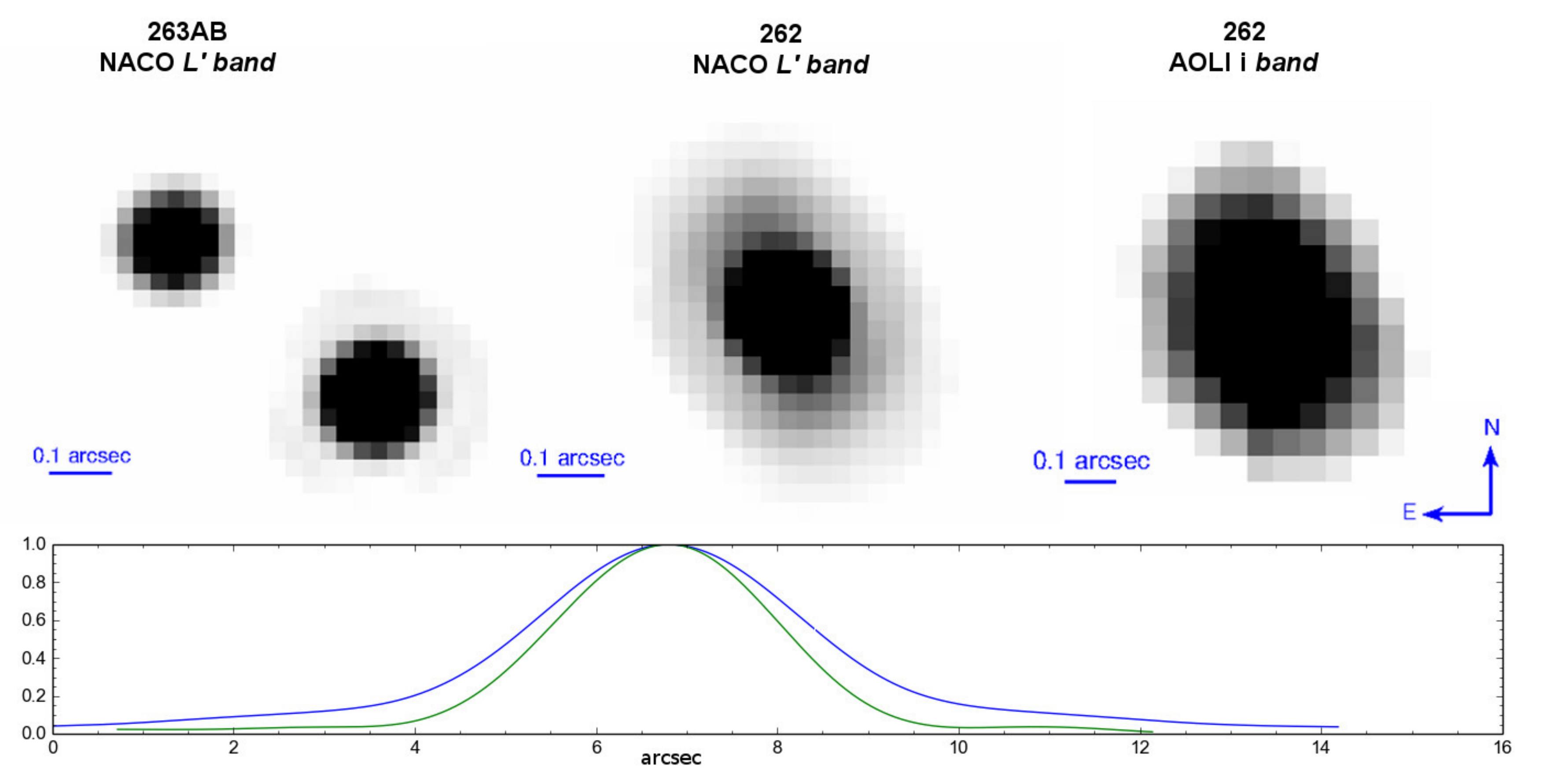}
	\caption{LkH$\alpha$ 263AB (left) and 262 (middle) as seen by NACO, taken under program 074.C-0789(A), PI S. Correia at the UT4-Yepun telescope (VLT) in the  \textit{L'} band, and LkH$\alpha$ 262 by AOLI (right) in the \textit{i'} band. The two-dimensional cut of the 262 (blue) and 263A (green) profiles along the line of maximum elongation for the NACO images are shown in the bottom diagram. Note the existence of an elongation in both the IR and optical images of LkH$\alpha$ 262 and the wider PSF of LkH$\alpha$ 262 relative to the components of 263.}
	\label{fig:comparanaco}
\end{figure}

When this was recognized, we carried out an extensive archival search with the aim of finding high-resolution images of LkH$\alpha$ 262 and we found observations obtained with NACO by S.Correia at the Very Large Telescope (VLT) in the \textit{ 2.17, 2.12, and 1.64} $\mu$m bands \citep{2006A&A...459..909C} and in the \textit{L'} band (unpublished). In these observations, we have marginally detected the  possible companion to LkH$\alpha$ 262 in the \textit{L'} band (see Figure \ref{fig:comparanaco}). Although there are other infrared AO observations, i.e.\citet{2002ApJ...571L..51J}, the presence of this companion had not been previously reported. 

The characteristics of object 262B remain to be precisely determined; however, it seems that both 262A and 262B would have a very similar flux in the \textit{L'} band  given the uniform elliptic shape. We have measured the ellipticity of  LkH$\alpha$ 262 to be 0.12, 0.22 and 0.14 in the images by Subaru, NACO and AOLI respectively. Our AOLI observations suggest that in the \textit{i'} band the NE object is slightly fainter, thus indicating a cooler and less massive object. We suggest that the SW object is the primary and propose it to be designated as LkH$\alpha$ 262A.

We have analysed the error sources for the relative astrometric position between LkH$\alpha$ 262 and 263AB for all the measurements given in Table \ref{table:astrometry2}:

- Centroid errors. These errors have been computed with the IRAF centroid routines applying a standard deviation of a series of independent measurements. Typical error is below 1/20 of the pixel size, i.e. 2 mas.

- Plate scale error. For each instrument we have computed plate scales from our own observations whenever it was possible.
In the data from other instruments we have used the reference values. Typical errors of plate scales are of order 1/100 of the
pixel size.  We have used Gaussian statistics to propagate the plate scale error  on the angular separation error. We  have combined the various sources of error, including the plate scale error and the centroid errors, to obtain the  final values of the errors in angular separation listed in Table \ref{table:astrometry2}. As a consistency check of these errors,  we note that the rms of the fit residuals of the angular separation measurements in Figure \ref{fig:pm} are 4 mas (optical) and 20 mas (near IR),  consistent with the smallest error bars in the respective plots. This small dispersion shows that the estimated individual error bars are  rather conservative.

- Atmospheric differential chromatic refraction (DCR effect). As we have conducted a multi-band imaging astrometric study, we have also computed the DCR effect. The dependence of the refractive index on the wavelength results in astrometric shifts of the centroids. Based on the equations given by \cite{1992AJ....103..638M} and \cite{1982PASP...94..715F}, and taking the approximation by \cite{2009AJ....138...19K}, we obtain an error introduced by the DCR of 5 mas, much smaller than the 130 mas shift between optical and NIR distance measurements between LkH$\alpha$ 262 and 263AB. 

The existence of a companion to 262 at a distance near the NACO diffraction limit in the $L'$ band, combined with the DCR effect, allows us to explain a difference of up to 70 mas between the distances measured in the optical and in NIR. The remaining 60 mas difference could be caused by the possible existence of discs in the LkH$\alpha$ 262/263 system. The existence of a disk in LkH$\alpha$ 262 was already claimed by Itoh et al. (2003). The presence of one or more additional discs could shorten the separation between LkH$\alpha$ 262 and 263AB at longer wavelengths, and help to explain the detected difference.

\begin{table*}
	\begin{minipage}{110mm}
		
		\caption{Binding energies for the LkH$\alpha$ 262-3 system.}
		\label{table:bind}
		
		\begin{tabular}{cccccc}
			\hline
			Primary     &  Secundary    & separation(s)  & $M_{1}$ & $M_{2}$ & $U^{*}_{g}$  \\
			&   &  [$10^{3}$ AU]   &  [$M_{\odot}$]  & [$M_{\odot}$])  & [$10^{33}J$] \\
			\hline
			\hline
			
			LkH$\alpha$ 263AB	& LkH$\alpha$ 262    & 4.281$\pm$0.003    & $0.8$ & $0.8$   &  -263.8       \\
			LkH$\alpha$ 263A	& LkH$\alpha$ 263B   & 0.112$\pm$0.008    & $0.4$ & $0.4$ &  -2521.5      \\
			LkH$\alpha$ 263AB   & LkH$\alpha$ 263C   & 1.131$\pm$0.005    & $0.8$ & $0.4$   &  -499.4      \\
			Sun	                & Neptune$^{a}$	     & 0.03               & 1.0  & 5.15  $\times10^{-5}$ & -3.0\\
			\hline
			
		\end{tabular}
		\medskip
		\small
		\\
		$^{a}$We have included Neptune as a comparison of a well-known bounded system.
	\end{minipage}
\end{table*}

\subsection{Comparison with stellar evolution models and dynamical considerations}
\label{age}

We compare the photometric properties of our objects with solar metallicity isochrones by using the Allard et al. (2011) PMS BT-settl models (CIFIST2011bc). These models are also described in \cite{{2014IAUS..299..271A}} and based on MHD simulations by Freytag et al. (2010). 

\begin{figure*}
	
	\includegraphics[width=0.8\linewidth]{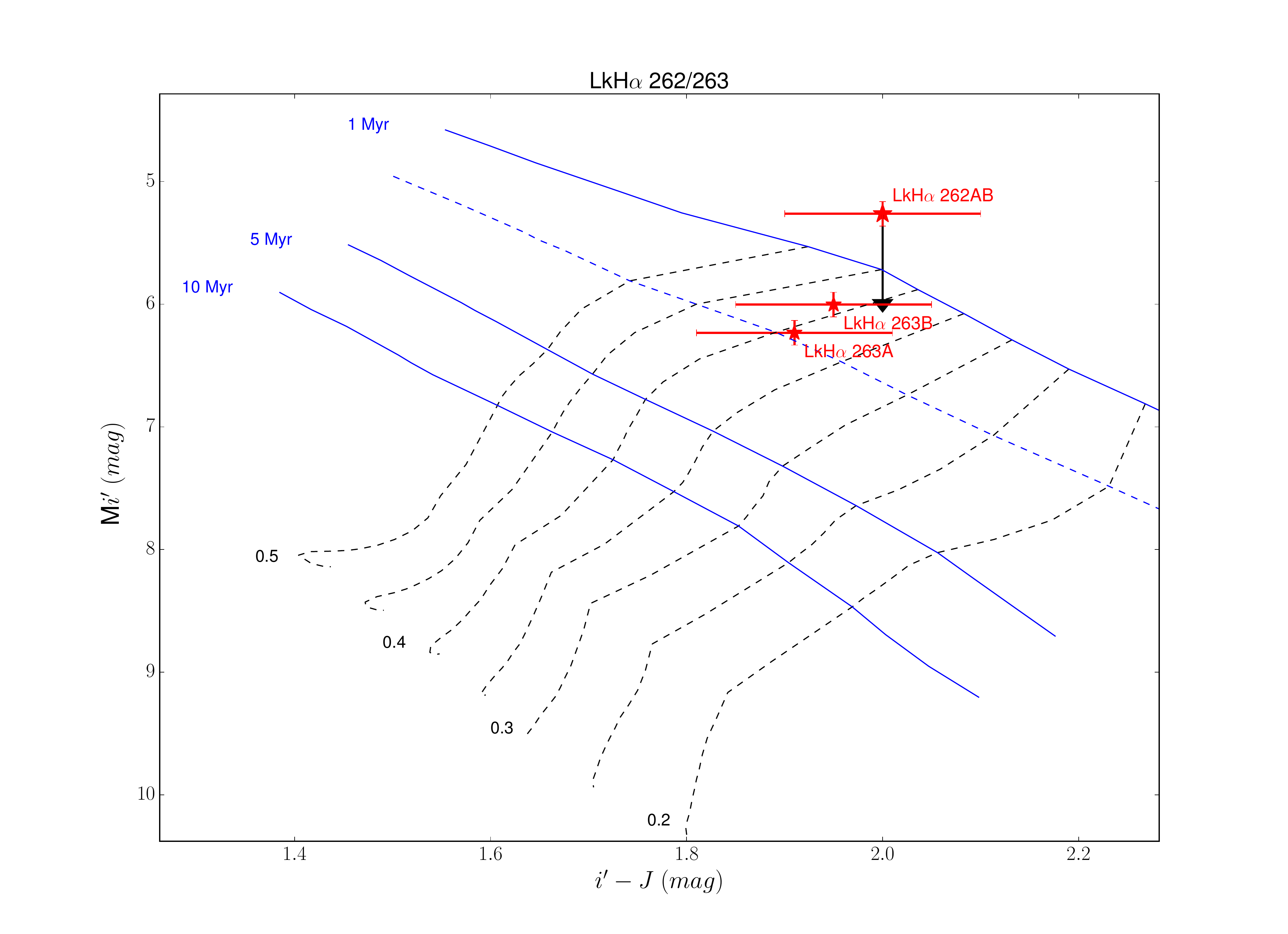}
	\caption{Location of LkH$\alpha$ 262/263 components in the absolute magnitude--colour diagram compared with PMS stellar evolution BT-Settl models from Allard et al., (2011). The dotted blue line represents the 2 Myr isochrone, whereas the dotted black lines represent the isomasses in solar mass units. Absolute magnitudes have been calculated for a distance of 275 pc, as given in \citet{2001ApJ...560..287L}. }
	\label{fig:isoc}
\end{figure*}

In the colour--magnitude diagram of Figure \ref{fig:isoc} it is shown that objects LkH$\alpha$ 263A and 263B are $\sim$ 0.4 solar masses with an age of a few Myr. We show with a black arrow the plausible position of LkH$\alpha$ 262A or 262B under the hypothesis of an equal brightness binary. The location indicated by the arrow shows that this is fully compatible with the properties of the LkH$\alpha$ 263 A and B components.

Meanwhile, LkH$\alpha$ 263C, with a central M0 star \citep{2002ApJ...571L..51J}, is located far  from the coeval components in the diagram. This is a result of the presence of an optically thick edge-on disc around it, which extinguishes the flux emerging from the star and also possibly redistributes it, enhancing the optical with respect to the infrared by about 1 magnitude.

We assume that the LkH$\alpha$ 262/263 system lies 275 pc away from Earth. Component B appears at 0.40 arcsec from component A, meaning a physical separation of 112 AU. For these two components, the change of position angle between the observations made by \citet{2002ApJ...571L..51J} and AOLI is 0.80$\pm$0.15 degrees in  the 12.85-year period of time between both measurements. If a circular orbit is assumed, the orbital period would be around 1200 years. 

For the time baseline between observations, the theoretical change in the angle should be around 3 degrees. However, we got a much lower value, which suggests that the orbit is not perpendicular to our line of sight, with a probable orbital inclination about $\pm12$ degrees, i.e.\ we see it almost edge-on. If this is the case, the orbit of 263AB would be coplanar with the  edge-on disc around component C. 

Component C lies 4 arcsec away from component A, which means a physical separation of 1000 AU. The position angle change appears to be 1.0$\pm$0.4 degrees in a time interval of 13 years. However, these measurements have been compared between the NIR and the optical, and the central star is surrounded by a wide optically thick disc; thus, the centroid may vary with wavelength. Additional measurements in the same bands are required to try to find evidence for the orbital motion of component C, which is expected to be of order 0.01 degrees per year.

A separation of 15 arcsec ($\sim$4000 AU at that distance) does not represent an impediment for the LkH$\alpha$ 262-263 system to be bounded, see \citet{2011ApJS..192....2S}. We have computed the binding energies for all the system components, see Table \ref{table:bind}. Wide binary systems have very low gravitational potential energy ruled by $U_{g}=-GM_{1}M_{2}/r$. We have used the separation between components (s) to obtain the gravitational potential (binding) energy listed in Table \ref{table:bind}.

\section{Conclusions}

\label{sec:conclu}

The observations performed with AOLI in the \textit{i'} band allowed us to  resolve spatially the  LkH$\alpha$ 263 components in the optical band. By means of these observations and through comparing the colours of the components in the LkH$\alpha$ 263AB pair, as well as using reference values for M-type stars, we have been able to give a first approximation to their spectral classification, component A being an M0--M1 star and component B an M1--M2 star. The AB pair is accompanied by the spectroscopically studied M0 component C.  

Our study revealed that the LkH$\alpha$ 263AB pair undergoes a change in the relative position angle of 0.06$\pm$0.01 degrees per year while their separation has remained unaltered to within the precision of the measurements. Thus, we can state that they are gravitationally bound. We also find a probably edge-on orbit for this pair that would consequently be close to coplanar with the disc in component C.

We have also studied the separation of LkH$\alpha$ 262 and LkH$\alpha$ 263 as a function of time and find that it changed less than 3 mas over the 10 year timespan, being  15.589$\pm$0.006 arcsec at the time of the AOLI observation. The differential motion has been constrained to be less than 1 mas/yr, about 10 times smaller than the typical proper motions of stars at the distance of the system.This indicates that the system is very likely gravitationally bound.  

We have also measured a discrepancy in the separation between LkH$\alpha$ 262 and 263 in the optical with respect to the infrared and noticed that the PSF of LkH$\alpha$ 262 appears to be elongated in the same direction in both optical and infrared images. This could be explained by the presence of a cooler companion of LkH$\alpha$ 262 at less than 0.15 arcsec, turning the system into a quintuple one.

All the components in the LkH$\alpha$ 262/263 system are young (less than 2 Myr), low mass (0.4 M$\odot$) PMS stars.

\section*{Acknowledgments}

This paper is based on observations made with the William Herschel Telescope operated on the island of La Palma by the Isaac Newton Group in the Spanish Observatorio del Roque de los Muchachos of the Instituto de Astrof\'isica de Canarias, with the Nordic Optical Telescope operated  by the Nordic Optical Telescope Scientific Association in the Spanish Observatorio del Roque de los Muchachos of the Instituto de Astrof\'isica de Canarias and with the IAC80 and Carlos S\'anchez Telescopes operated on the island of Tenerife by the Instituto de Astrof\'isica de Canarias in the Spanish Observatorio del Teide. We are very grateful to the ING staff and the IAC Support Astronomers Group for their efforts. 

The data presented here were obtained in part with ALFOSC, which is provided by the Instituto de Astrofisica de Andalucia (IAA) under a joint agreement with the University of Copenhagen and NOTSA.

This research has made use of the Washington Double Star Catalogue maintained at the U.S. Naval Observatory.

Some of the data presented in this paper were obtained from the Mikulski Archive for Space Telescopes (MAST). STScI is operated by the Association of Universities for Research in Astronomy, Inc., under NASA contract NAS5-26555. Support for MAST for non-HST data is provided by the NASA Office of Space Science via grant NNX13AC07G and by other grants and contracts.

A.P.G. and A.D.S. have been supported by Project No. 15345/PI/10 from the Fundaci\'on S\'eneca and MINECO under the grant AYA2011-29024.

The authors thank the referee for his/her helpful comments.











\bsp	
\label{lastpage}
\end{document}